\documentclass[prd,reprint,showpacs,showkeys]{revtex4-1}
\usepackage{amsfonts}
\usepackage{amssymb}
\usepackage{amsmath}
\usepackage{graphicx}
\usepackage[font={footnotesize,it}]{caption}

\begin{document}

\preprint{}
\title{Non-abelian magnetic black strings versus black holes}
\author{S. Habib Mazharimousavi}
\email{habib.mazhari@emu.edu.tr}
\author{M. Halilsoy}
\email{mustafa.halilsoy@emu.edu.tr}
\affiliation{Department of Physics, Eastern Mediterranean University, G. Magusa, North
Cyprus, Mersin 10 - Turkey.}

\begin{abstract}
We present $d+1-$dimensional pure magnetic Yang-Mills (YM) black strings (or 
$1-$branes) induced by the $d-$dimensional Einstein-Yang-Mills-Dilaton black
holes. Born-Infeld version of the YM field makes our starting point which
goes to the standard YM field through a limiting procedure. The lifting from
black holes to black strings, (with less number of fields) is by adding an
extra, compact coordinate. This amounts to the change of horizon topology
from $S^{d-2}$ to a product structure. Our black string in $5-$dimensions is
a rather special one, with uniform Hawking temperature and
non-asymptotically flat structure. As the YM charge becomes large the string
gets thinner to tend into a breaking point and transform into a $4-$%
dimensional black hole.
\end{abstract}

\pacs{PACS number}
\maketitle

\section{Introduction}

A black string, as a black object \cite{1,2,3,4,5,6} is the simplest
extension, of a black hole, with the addition of a compact, killing
coordinate. This definition does not include the non-uniform black strings
in which the metric functions are explicitly dependent on the
extra-coordinate \cite{7,8,9}. More than one such additions makes the class
of black $p-$branes whose first member, $1-$brane is a black string. The
addition of an extra, compact coordinate does affect the Komar mass and
total charge of the system, since it arises in the total volume element. As
the energy involves the compact dimension that means, it arises also in the
thermodynamical functions, heat capacity, phase transitions and other
properties. Divergence in these thermodynamical functions is known to create
instability in the system. Decay of a black string into a black hole (or
vice versa) is one of the possible scenarios that may take place in this
context. Depending on the critical mass of a black string a possible
instability was formulated through perturbation analysis by Gregory and
Laflamme (GL), which is known as the GL instability \cite{10}. It was
further argued that this kind of instability threatens all black objects,
not only black strings with a single compact dimension. In the presence of a
magnetic charge and dilaton a black string may be stable when confined
within a range of compact dimension \cite{11}. A similar instability was
demonstrated also for an electrically charged black string \cite{12}.

In this paper our purpose is to construct generically non-asymptotically
flat (NAF) and non-spherically symmetric static black strings in higher
dimensions from pure magnetically charged Yang-Mills (YM) fields and
dilatons. To this end we employ certain classes of
Einstein-Yang-Mills-Born-Infeld-Dilaton (EYMBID) solutions that we had
obtained before \cite{13}. Black strings constructed from abelian Maxwell
fields were known \cite{14}, so we wish to carry the construction to cover
the non-abelian YM fields. Our method of introducing the YM fields makes use
of the Wu-Yang ansatz \cite{15} which was extended to all higher dimensions 
\cite{16,17,18}. When restricted to $d=4$, our YM field becomes almost same
with the abelian Maxwell field \cite{19}, therefore in our procedure higher (%
$d\geq 5$) dimensions become significant. We had found also the YM version
of the Born-Infeld (BI) electrodynamics which was devised originally to
remove divergences in classical electromagnetism. In particular limits we
can remove the BI contribution to retain the usual YM-dilaton coupling in
our model. The original Kaluza-Klein (KK) theory was to employ less (or no)
fields other than gravity and to derive physical fields through a
dimensional reduction. Thus, one can generate dilaton and electromagnetic
fields from pure gravity by proper reduction procedure. Likewise, by lifting
the spacetime to higher dimensions we dissolve some of the fields and
simplify our action to the extreme of pure geometrical one. In addition, in
this study we compare the entropies of black strings and black holes of same
dimensionality to see which one is favorable. Since the natural flow is from
a smaller to a larger entropy state this at least gives us an analytic
method to find which state is more stable or not, to decay into the other.
The possibility of decay into a group of naked singularities has been
considered recently \cite{20}.

\section{EYMBID theory in brief}

In this section we give a short review on the black hole solution in EYMBID
theory found previously in \cite{13}. For details one may consider Ref. \cite%
{10} as well. The $d-$dimensional action in EYMBID theory is given by 
\begin{equation}
I=\frac{1}{16\pi G_{\left( d\right) }}\int\nolimits_{\mathcal{M}}d^{d}x\sqrt{%
-\bar{g}}\left( -\bar{R}+\frac{4}{d-2}\left( \mathbf{\bar{\nabla}}\psi
\right) ^{2}-L\left( \mathbf{\bar{F}},\psi \right) \right) ,
\end{equation}%
where $\bar{R}$ is the Ricci scalar, $\psi $ is the dilaton and YM 2-form
field is defined by \cite{16,17,18} 
\begin{equation}
\mathbf{\bar{F}}^{\left( a\right) }=\mathbf{d\bar{A}}^{\left( a\right) }+%
\frac{1}{2\sigma }C_{\left( b\right) \left( c\right) }^{\left( a\right) }%
\mathbf{\bar{A}}^{\left( b\right) }\wedge \mathbf{\bar{A}}^{\left( c\right) }
\end{equation}%
with structure constants $C_{\left( b\right) \left( c\right) }^{\left(
a\right) }.$ Our choice of YM potential $\mathbf{\bar{A}}^{\left( a\right) }$
follows from the higher dimensional Wu-Yang ansatz \cite{15,16,17,18} and
the coupling constant $\sigma $ is expressed in terms of the YM charge. The
Lagrangian $L\left( \mathbf{\bar{F}},\psi \right) $ is chosen as 
\begin{multline}
L\left( \mathbf{\bar{F}},\psi \right) =4\beta ^{2}e^{4\alpha \psi /\left(
d-2\right) }\times \\
\left( 1-\sqrt{1+\frac{\mathbf{Tr}(\bar{F}_{\lambda \sigma }^{\left(
a\right) }\bar{F}^{\left( a\right) \lambda \sigma })e^{-8\alpha \psi /\left(
d-2\right) }}{2\beta ^{2}}}\right)
\end{multline}%
in which $\alpha $ and $\beta $ are the dilaton and BI parameters,
respectively. For brevity, let us introduce the following abbreviation%
\begin{equation}
\mathcal{L}\left( X\right) =1-\sqrt{1+X},
\end{equation}%
in which 
\begin{equation*}
\text{ \ \ }X=\frac{\mathbf{Tr}(\bar{F}_{\lambda \sigma }^{\left( a\right) }%
\bar{F}^{\left( a\right) \lambda \sigma })e^{-8\alpha \psi /\left(
d-2\right) }}{2\beta ^{2}},\text{ \ \ }\mathbf{Tr}(.)=\overset{\left(
d-1\right) (d-2)/2}{\underset{a=1}{\sum }\left( .\right) .}
\end{equation*}%
Our choice for the metric ansatz is 
\begin{equation}
d\bar{s}^{2}=-\bar{f}\left( r\right) dt^{2}+\frac{dr^{2}}{\bar{f}\left(
r\right) }+\bar{h}\left( r\right) ^{2}d\Omega _{\left( d-2\right) }^{2},
\end{equation}%
in which $\bar{f}\left( r\right) $ and $\bar{h}\left( r\right) $ stand for
functions of $r$ and $d\Omega _{\left( d-2\right) }^{2}$ is the $\left(
d-2\right) -$dimensional unit spherical line element. Variation of (1) with
respect to $\bar{g}_{\mu \nu }$ and $\psi $ yields%
\begin{multline}
\bar{R}_{\mu \nu }=\frac{4}{d-2}\mathbf{\bar{\partial}}_{\mu }\psi \bar{%
\partial}_{\nu }\psi - \\
4e^{-4\alpha \psi /\left( d-2\right) }\left( \mathbf{Tr}\left( \bar{F}_{\mu
\lambda }^{\left( a\right) }\bar{F}_{\nu }^{\left( a\right) \ \lambda
}\right) \partial _{X}\mathcal{L}\left( X\right) \right) + \\
\frac{4\beta ^{2}}{d-2}e^{4\alpha \psi /\left( d-2\right) }\mathcal{K}\left(
X\right) \bar{g}_{\mu \nu },
\end{multline}%
and%
\begin{equation}
\bar{\nabla}^{2}\psi =2\alpha \beta ^{2}e^{4\alpha \psi /\left( d-2\right) }%
\mathcal{K}\left( X\right) ,
\end{equation}%
in which we have further abbreviated 
\begin{equation}
\mathcal{K}\left( X\right) =2X\partial _{X}\mathcal{L}\left( X\right) -%
\mathcal{L}\left( X\right)
\end{equation}%
where $\partial _{X}\mathcal{L}\left( X\right) =\frac{\partial \mathcal{L}%
\left( X\right) }{\partial X}$.\ The YM equations take the form 
\begin{multline}
\mathbf{d}\left( e^{-4\alpha \psi /\left( d-2\right) \star }\mathbf{\bar{F}}%
^{\left( a\right) }\right) + \\
\frac{1}{\sigma }C_{\left( b\right) \left( c\right) }^{\left( a\right)
}e^{-4\alpha \psi /\left( d-2\right) }\mathbf{\bar{A}}^{\left( b\right)
}\wedge ^{\star }\mathbf{\bar{F}}^{\left( c\right) }=0
\end{multline}%
where the hodge star $^{\star }$ implies duality. It can readily be seen
that for $\beta \rightarrow 0$ (i.e. no YM field) Eq.s (6)-(7) reduce to%
\begin{equation}
\bar{R}_{\mu \nu }=\frac{4}{d-2}\mathbf{\bar{\partial}}_{\mu }\psi \bar{%
\partial}_{\nu }\psi ,
\end{equation}%
and%
\begin{equation}
\bar{\nabla}^{2}\psi =0\text{\ }
\end{equation}%
which refer simply to the gravity coupled with a massless scalar field. The
limit $\beta \rightarrow \infty $ removes BI theory and we obtain simply
EYMD theory, which will be the subject of Sec. III.

Exact solution for the metric (5) and dilaton are given by \cite{13}%
\begin{equation}
\bar{f}\left( r\right) =\Xi \left( 1-\left( \frac{r_{+}}{r}\right) ^{\frac{%
\left( d-3\right) \alpha ^{2}+1}{\alpha ^{2}+1}}\right) r^{\frac{2}{\alpha
^{2}+1}},
\end{equation}%
\begin{equation}
\bar{h}\left( r\right) =Ae^{-2\alpha \psi /\left( d-2\right) },\text{ \ \ (}%
A=\text{constant)}
\end{equation}%
and 
\begin{equation}
\psi =-\frac{\left( d-2\right) }{2}\frac{\alpha \ln r}{\alpha ^{2}+1},
\end{equation}%
in which 
\begin{equation}
\Xi =-\frac{4\beta ^{2}\left( \alpha ^{2}+1\right) ^{2}\mathcal{K}\left(
X\right) }{\left( d-2\right) \left( \left( d-3\right) \alpha ^{2}+1\right) }
\end{equation}%
and%
\begin{equation}
r_{+}=\left( \frac{4\left( \alpha ^{2}+1\right) M_{QL}}{\left( d-2\right)
\Xi \alpha ^{2}A^{d-2}}\right) .
\end{equation}%
Here $M_{QL}$ stands for the Quasi-local mass \cite{21} and the constants
satisfy the constraint condition%
\begin{multline}
4\mathcal{K}\left( X\right) \beta ^{2}A^{4}\left( \alpha ^{2}-1\right) + \\
\left( d-2\right) \left( d-3\right) \left( 4Q^{2}\partial _{X}\mathcal{L}%
+A^{2}\right) =0.
\end{multline}%
It is of utmost importance to comment at this point, with reference to \cite%
{13}, that the solution for $\psi $ and the invariant of YM field $\mathbf{Tr%
}(\bar{F}_{\lambda \sigma }^{\left( a\right) }\bar{F}^{\left( a\right)
\lambda \sigma })$ just compensates each other to yield $X=const.,$ namely 
\begin{equation}
X=\frac{\left( d-2\right) \left( d-3\right) Q^{2}}{2\beta ^{2}A^{4}}.
\end{equation}%
As a result we have $\mathcal{L}\left( X\right) =const.,$ $\mathcal{K}\left(
X\right) =const.$ and $\Xi =const.,$ so that the foregoing expressions
become meaningful.

\section{EYMBI black string}

Now let's consider a $\left( d+1\right) -$dimensional space time as%
\begin{equation}
ds^{2}=e^{-b\psi }d\bar{s}^{2}+e^{b\left( d-2\right) \psi }dz^{2}
\end{equation}%
in which $b=$constant, $d\bar{s}^{2}$ is our line element (5) and $z$ is an
additional, compact Killing coordinate ($0<z\leq L$). Our new action
modifies into%
\begin{multline}
I=\frac{-1}{16\pi G_{\left( d+1\right) }}\int_{0}^{L}dz\int d^{d}x\sqrt{-g}%
\times \\
\left( R+4\beta ^{2}\left( 1-\sqrt{1+\frac{\mathbf{Tr}(F_{\lambda \sigma
}^{\left( a\right) }F^{\left( a\right) \lambda \sigma })}{2\beta ^{2}}}%
\right) \right) ,
\end{multline}%
where the $d+1$ and $d-$dimensional quantities are related as follow%
\begin{equation}
\sqrt{-g}=e^{-b\psi }\sqrt{-\bar{g}},
\end{equation}%
\begin{equation}
R=e^{b\psi }\left[ \bar{R}+b\left\{ \bar{\nabla}^{2}\psi -\frac{\left(
d-1\right) \left( d-2\right) }{4}b\left( \bar{\nabla}\psi \right)
^{2}\right\} \right] ,
\end{equation}%
\begin{equation}
\mathbf{A}^{(a)}=\mathbf{\bar{A}}^{(a)},
\end{equation}%
and%
\begin{equation}
\mathcal{F}=e^{2b\psi }\mathcal{\bar{F}},
\end{equation}%
in which%
\begin{equation}
\mathcal{F}=F_{\lambda \sigma }^{\left( a\right) }F^{\left( a\right) \lambda
\sigma },\text{ }\mathcal{\bar{F}}=\bar{F}_{\lambda \sigma }^{\left(
a\right) }\bar{F}^{\left( a\right) \lambda \sigma }\text{.}
\end{equation}%
It is a straightforward calculation to show that while $\bar{F}^{(a)}$
satisfies the YM equations in $d-$dimensions $F^{(a)}=\bar{F}^{(a)}$
satisfies the YM equations in $d+1-$dimensions. By substitutions directly
into (20) and integrating over the compact coordinate (i.e. $%
\int\limits_{0}^{L}dz=L$.) we arrive at the action 
\begin{multline}
I=\frac{-1}{16\pi G_{\left( d\right) }}\int d^{d}x\sqrt{-\bar{g}}\times
\left( \bar{R}-\frac{4\left( \mathbf{\bar{\nabla}}\psi \right) ^{2}}{d-2}%
+\right. \\
\left. 4\beta ^{2}e^{\frac{4\psi }{\left( d-2\right) \sqrt{d-1}}}\left( 1-%
\sqrt{1+\frac{\mathcal{\bar{F}}e^{\frac{-8\psi }{\left( d-2\right) \sqrt{d-1}%
}}}{2\beta ^{2}}}\right) \right) ,
\end{multline}%
in which the Newton's gravitational constant $G_{\left( d\right) }$ is
related to its 4d version by $G_{\left( d\right) }=G_{\left( 4\right)
}L^{d-4}$. This is nothing but the $d-$dimensional action for the choice $%
\alpha =\frac{1}{\sqrt{d-1}}$ and $b=-\frac{4}{\left( d-2\right) \sqrt{d-1}}$
in the EYMBID theory. Now, by expressing the $d+1-$dimensional black string
metric in the form%
\begin{multline}
ds^{2}=-f_{1}\left( r\right) dt^{2}+\frac{dr^{2}}{f_{2}\left( r\right) }+ \\
f_{3}\left( r\right) dz^{2}+f_{4}\left( r\right) d\Omega _{\left( d-2\right)
}^{2}
\end{multline}%
in which $f_{i}\left( r\right) $ are functions of $r$ only serves to
construct the latter from the known solution of EYMBID black hole solution
in $d-$dimensions. The routine identification of the metric functions go as
follows%
\begin{equation}
f_{1}\left( r\right) =e^{-b\psi }\bar{f}\left( r\right) ,
\end{equation}%
\begin{equation}
f_{2}\left( r\right) =e^{b\psi }\bar{f}\left( r\right) ,
\end{equation}%
\begin{equation}
f_{3}\left( r\right) =e^{b\left( d-2\right) \psi },
\end{equation}%
and%
\begin{equation}
f_{4}\left( r\right) =e^{-b\psi }\bar{h}\left( r\right) ^{2}
\end{equation}%
where $\psi $, $\bar{f}\left( r\right) $ and $\bar{h}\left( r\right) $ are
given in (12)-(15). Let us note that in lifting of the solution to the $d+1-$%
dimensional string solution the dilatonic component already disappeared. The
EYMBI black string solution given in (27) has a horizon given by $%
f_{1}\left( r_{+}\right) =0,$ which is given by (16). The entropy is defined
as 
\begin{equation}
S=\frac{A_{H}}{4G_{\left( d+1\right) }},
\end{equation}%
in which 
\begin{equation}
A_{H}=\frac{2\pi ^{\frac{d-1}{2}}L}{\Gamma \left( \frac{d-1}{2}\right) }%
f_{4}\left( r_{+}\right) ^{\frac{d-2}{2}}\sqrt{f_{3}\left( r_{+}\right) }
\end{equation}%
and%
\begin{equation}
G_{\left( d+1\right) }=G_{\left( 4\right) }L^{d-3}=\frac{L^{d-3}}{16\pi }.
\end{equation}%
Here we set $16\pi G_{\left( 4\right) }=1$ and therefore%
\begin{equation}
S=\frac{8}{L^{d-4}}\frac{\pi ^{\frac{d+1}{2}}f_{4}\left( r_{+}\right) ^{%
\frac{d-2}{2}}\sqrt{f_{3}\left( r_{+}\right) }}{\Gamma \left( \frac{d-1}{2}%
\right) }.
\end{equation}%
Explicitly one finds%
\begin{equation}
f_{1}\left( r\right) =\Xi \left( r^{2\left( d-2\right) /d}-r_{+}^{2\left(
d-2\right) /d}\right) ,
\end{equation}%
\begin{equation}
f_{2}\left( r\right) =\Xi r^{4/d}\left( r^{2\left( d-2\right)
/d}-r_{+}^{2\left( d-2\right) /d}\right) ,
\end{equation}%
and%
\begin{equation}
f_{3}\left( r\right) =r^{\frac{2\left( d-2\right) }{d}},\text{ \ \ }%
f_{4}\left( r\right) =A^{2}=const..
\end{equation}%
A substitution in (35) implies%
\begin{equation}
S=\frac{8}{L^{d-4}}\frac{\pi ^{\frac{d+1}{2}}}{\Gamma \left( \frac{d-1}{2}%
\right) }\xi ^{d-2}r_{+}^{\frac{\left( d-2\right) }{d}}.
\end{equation}%
Now we calculate the Hawking temperature 
\begin{equation}
T_{H}=\frac{1}{4\pi }f_{1}^{\prime }\left( r_{+}\right) =\frac{\Xi \left(
d-2\right) }{2d\pi }r_{+}^{\frac{d-4}{d}}
\end{equation}%
and the specific heat capacity of the black string%
\begin{equation}
C_{q}=T_{H}\left( \frac{\partial S}{\partial T_{H}}\right) _{q}=\frac{d-2}{%
d-4}S
\end{equation}%
which clearly diverges for $d=4.$ This is expected as the Hawking
temperature in (40) for $d=4$ is constant with respect to $r_{+}$.

\section{EYM black string}

From the action (1) it can be shown easily that 
\begin{equation}
\lim_{\beta \rightarrow \infty }\text{ }L\left( \mathbf{\bar{F}},\psi
\right) =-e^{-4\alpha \psi /\left( d-2\right) }\mathcal{\bar{F}}
\end{equation}%
which is the dilaton-YM coupling Lagrangian and accordingly the resulting
field equations reduce to%
\begin{multline}
\bar{R}_{\mu }^{\nu }=\frac{4}{d-2}\delta _{\mu }^{\nu }\left( \mathbf{\bar{%
\nabla}}\psi \right) ^{2}+ \\
2e^{-4\alpha \psi /\left( d-2\right) }\left[ \mathbf{Tr}\left( \bar{F}_{\mu
\lambda }^{\left( a\right) }\bar{F}^{\left( a\right) \nu \lambda }\right) -%
\frac{1}{2\left( d-2\right) }\mathcal{\bar{F}}\delta _{\mu }^{\nu }\right] ,
\end{multline}%
and 
\begin{equation}
\bar{\nabla}^{2}\psi =-\frac{1}{2}\alpha e^{-4\alpha \psi /\left( d-2\right)
}\mathcal{\bar{F}}.
\end{equation}%
With the invariant YM equation and same ansatz metric as (5) we obtain the
solution%
\begin{equation}
\psi =-\frac{\left( d-2\right) }{2}\frac{\alpha \ln r}{\alpha ^{2}+1}
\end{equation}%
\begin{equation}
\bar{f}\left( r\right) =\Xi \left( 1-\left( \frac{r_{+}}{r}\right) ^{\frac{%
\left( d-3\right) \alpha ^{2}+1}{\alpha ^{2}+1}}\right) r^{\frac{2}{\alpha
^{2}+1}},
\end{equation}%
and%
\begin{equation}
\bar{h}\left( r\right) =\xi e^{-2\alpha \psi /\left( d-2\right) },
\end{equation}%
in which%
\begin{equation}
\Xi =\frac{\left( d-3\right) }{\left( \left( d-3\right) \alpha ^{2}+1\right)
Q^{2}},\text{ }\xi ^{2}=Q^{2}\left( \alpha ^{2}+1\right) .
\end{equation}%
The corresponding EYM black string metric is same as (27), and upon setting $%
\alpha =\frac{1}{\sqrt{d-1}}$ and $b=-\frac{4}{\left( d-2\right) \sqrt{d-1}}$
the $d-$dimensional action casts into 
\begin{multline}
I=\frac{-1}{16\pi G_{\left( d\right) }}\int\nolimits_{\mathcal{M}}d^{d}x%
\sqrt{-\bar{g}}\times  \\
\left( \bar{R}-\frac{4}{d-2}\left( \mathbf{\bar{\nabla}}\psi \right)
^{2}-e^{-4\alpha \psi /\left( d-2\right) }\mathcal{\bar{F}}\right) .
\end{multline}%
By choosing the $(d+1)-$line element as in (27) we identify the EYM black
string metric functions easily. Here also it is remarkable that the physical
properties of the black string is identical with those given in Eq.s
(39)-(41). Specifically we concentrate on the entropy of the EYM black
string which can be written as%
\begin{equation}
S_{BS}=\frac{8}{L^{d-4}}\frac{\pi ^{\frac{d+1}{2}}}{\Gamma \left( \frac{d-1}{%
2}\right) }A^{d-2}r_{+}^{\frac{\left( d-2\right) }{d}}.
\end{equation}%
Our aim is to compare the entropy of $d+1-$dimensional EYMBS with the
entropy of the $d+1-$dimensional EYM black hole which was found recently in 
\cite{18}. The spherically and static line element of the solution reported
in \cite{18} is given by%
\begin{equation}
ds_{BH}^{2}=-f\left( r\right) dt^{2}+\frac{1}{f\left( r\right) }%
dr^{2}+r^{2}d\Omega _{d-1}^{2}
\end{equation}%
where%
\begin{equation}
f\left( r\right) =\left\{ 
\begin{array}{ccc}
1-\frac{m}{r^{2}}-\frac{2Q^{2}}{r^{2}}\ln r, &  & d=4 \\ 
1-\frac{m}{r^{d-3}}-\frac{\left( d-2\right) Q^{2}}{\left( d-4\right) r^{2}},
&  & d\geq 5%
\end{array}%
\right. 
\end{equation}%
in which $Q$ is the YM charge and $m$ is an integration constant related to
the mass of the black hole. (We note that $N$ in \cite{18} is $d+1$ in (51)
and (52) in this paper.) The entropy of the $d+1-$dimensional EYM black hole
(51) is given by%
\begin{equation}
S_{BH}=\frac{\pi ^{\frac{d+1}{2}}}{4\Gamma \left( \frac{d-1}{2}\right) }%
r_{h}^{d-1}
\end{equation}%
in which $r_{h}$ indicates the horizon of the black hole. Our final argument
is to define the micro-canonical equilibrium condition for the electric
black string as $S_{BS}>S_{BH}$. The corresponding critical curve is defined
as $S_{BS}=S_{BH}$ which describes a reversible transformation. Unlike the
usual approach we find the critical curve by equating the horizon of the BS
and BH i.e., $r_{+}=r_{h}.$ This leads to%
\begin{equation}
\frac{32A^{d-2}}{L^{d-4}}=r_{+}^{\frac{d\left( d-2\right) +2}{d}}
\end{equation}%
or equivalently%
\begin{equation}
\frac{32}{L^{d-4}}\left( \frac{Q^{2}d}{d-1}\right) ^{\frac{d-2}{2}}=r_{+}^{%
\frac{d\left( d-2\right) +2}{d}}.
\end{equation}

As an example to the foregoing analysis we obtain a particular $5D-$black
string as follows. We recall that in $d=4$, EYMD black hole solution
coincides with the NAF black hole solution in EMD gravity \cite{22,23}.
Thus, for $d=4$ we obtain%
\begin{equation}
\psi =-\frac{\alpha \ln r}{\alpha ^{2}+1},
\end{equation}%
\begin{equation}
\text{ }\bar{f}\left( r\right) =\Xi \left( 1-\left( \frac{r_{+}}{r}\right)
\right) r^{\frac{3}{2}},\text{ }
\end{equation}%
\begin{equation}
\bar{h}\left( r\right) =\xi r^{1/4},
\end{equation}%
in which $\Xi =\frac{3}{4Q^{2}},$ $\xi ^{2}=\frac{4}{3}Q^{2},$ $\alpha =%
\frac{1}{\sqrt{3}},$ $b=-\frac{2}{\sqrt{3}}$ and%
\begin{equation}
e^{-b\psi }=r^{-1/2}.
\end{equation}%
\ Now from the identification (28)-(31) the $5D-$black string metric follows 
\begin{multline}
ds^{2}=-\frac{3\left( r-r_{+}\right) }{4Q^{2}}dt^{2}+ \\
\frac{4Q^{2}}{3r\left( r-r_{+}\right) }dr^{2}+rdz^{2}+\frac{4Q^{2}}{3}%
d\Omega _{2}^{2}.
\end{multline}%
This line element has the Ricci ($R$) and Kretchmann ($K$) scalars as%
\begin{equation}
R=\frac{3}{8Q^{2}}
\end{equation}%
and%
\begin{equation}
K=\frac{171}{64Q^{2}}
\end{equation}%
which reveal the regularity of the string metric. For large YM charge it is
observed that these invariants get smaller. The resulting action is
expressed by 
\begin{equation}
I=\frac{-1}{16\pi G_{\left( 5\right) }}\int_{0}^{L}dz\int d^{4}x\sqrt{-g}%
\left( R-\mathcal{F}\right)
\end{equation}%
in which $\mathcal{F}$ is the Maxwell invariant without dilaton. Here we
also calculate the Hawking temperature 
\begin{equation}
T_{H}=\frac{1}{4\pi }f_{1}^{\prime }\left( r_{+}\right) =\frac{3}{16\pi Q^{2}%
}
\end{equation}%
which is constant.

\section{CONCLUSION}

Black strings in $(d+1)-$dimensions are constructed from $d-$dimensional
black holes in the presence of non-abelian YM fields. Particular black
strings are obtained from the EYMD black holes. Dilatonic coupling is strong
enough to convert spacetime into a NAF one. We go one more step ahead to
consider BI coupling out of the YM field which makes the overall model
highly non linear. In the limit ($\beta \rightarrow \infty $) of BI we
recover the standard YM theory coupled with the dilaton. Our explicite $5-$%
dimensional black string (58) has a unique structure: it admits a uniform
Hawking temperature $T_{H}$, irrespective of both the horizon radius and
quasi local mass. This implies that any thermodynamical stability analysis
on $T_{H}$ fails to work. We didn't investigate the GL type instability for
our black string, but it is our belief that it remains intact. For the
particular dimension $d=4$ we obtain a diverging specific heat $%
C_{q}\rightarrow \infty $ which signals a phase transition from a black
string to other structures. The equality $S_{BH}=S_{BS}$, however, suggests
a reversible transformation between the BS and BH. Any entropy dominance
between the two will support the existence of one relative to the other.

\end{document}